\documentstyle[preprint,eqsecnum,epsf,aps,floats,tighten]{revtex}
\def\met{\mbox{${\hbox{$E$\kern-0.6em\lower-.1ex\hbox{/}}}_T$}} 
\def\D0{D\O}                            
\def\etal{{\sl et al.}}                 
\def\d0draft{}
\def\err#1#2#3 {{\it Erratum} {\bf#1},{\ #2} (19#3)}
\def\ib#1#2#3 {{\it ibid.} {\bf#1},{\ #2} (19#3)}
\def\nc#1#2#3 {Nuovo Cim. {\bf#1} ,#2(19#3)}
\def\nim#1#2#3 {Nucl. Instr. Meth. {\bf#1},{\ #2} (19#3)}
\def\np#1#2#3 {Nucl. Phys. {\bf#1},{\ #2} (19#3)}
\def\pl#1#2#3 {Phys. Lett. {\bf#1},{\ #2} (19#3)}
\def\prev#1#2#3 {Phys. Rev. {\bf#1},{\ #2} (19#3)}
\def\prl#1#2#3 {Phys. Rev. Lett. {\bf#1},{\ #2} (19#3)}
\def\rmp#1#2#3 {Rev. Mod. Phys. {\bf#1},{\ #2} (19#3)}
\def\zp#1#2#3 {Zeit. Phys. {\bf#1},{\ #2} (19#3)}

\def\ch0{\tilde{\chi^0_1}}

\begin{document}
\title{ \bf Search for R-parity Violating Supersymmetry in 
Dimuon and Four-Jets Channel}
%
\author{                                                                      
V.M.~Abazov,$^{23}$                                                           
B.~Abbott,$^{57}$                                                             
A.~Abdesselam,$^{11}$                                                         
M.~Abolins,$^{50}$                                                            
V.~Abramov,$^{26}$                                                            
B.S.~Acharya,$^{17}$                                                          
D.L.~Adams,$^{59}$                                                            
M.~Adams,$^{37}$                                                              
S.N.~Ahmed,$^{21}$                                                            
G.D.~Alexeev,$^{23}$
% ...
\\                                                                            
\vskip 0.30cm                                                                 
\centerline{(D\O\ Collaboration)}                                             
\vskip 0.30cm                                                                 
}                                                                             
\address{                                                                     
\centerline{$^{1}$Universidad de Buenos Aires, Buenos Aires, Argentina}       
\centerline{$^{2}$LAFEX, Centro Brasileiro de Pesquisas F{\'\i}sicas,         
                  Rio de Janeiro, Brazil}                                     
\centerline{$^{3}$Universidade do Estado do Rio de Janeiro,                   
                  Rio de Janeiro, Brazil}                                     
\centerline{$^{4}$Institute of High Energy Physics, Beijing,                  
                  People's Republic of China}                                 
\centerline{$^{5}$Universidad de los Andes, Bogot\'{a}, Colombia}             
\centerline{$^{6}$Charles University, Center for Particle Physics,            
                  Prague, Czech Republic}                                     
\centerline{$^{7}$Institute of Physics, Academy of Sciences, Center           
                  for Particle Physics, Prague, Czech Republic}               
\centerline{$^{8}$Universidad San Francisco de Quito, Quito, Ecuador}         
\centerline{$^{9}$Institut des Sciences Nucl\'eaires, IN2P3-CNRS,             
                  Universite de Grenoble 1, Grenoble, France}                 
\centerline{$^{10}$CPPM, IN2P3-CNRS, Universit\'e de la M\'editerran\'ee,     
                  Marseille, France}                                          
\centerline{$^{11}$Laboratoire de l'Acc\'el\'erateur Lin\'eaire,              
                  IN2P3-CNRS, Orsay, France}                                  
\centerline{$^{12}$LPNHE, Universit\'es Paris VI and VII, IN2P3-CNRS,         
                  Paris, France}                                              
\centerline{$^{13}$DAPNIA/Service de Physique des Particules, CEA, Saclay,    
                  France}                                                     
\centerline{$^{14}$Universit{\"a}t Mainz, Institut f{\"u}r Physik,            
                  Mainz, Germany}                                             
\centerline{$^{15}$Panjab University, Chandigarh, India}                      
\centerline{$^{16}$Delhi University, Delhi, India}                            
\centerline{$^{17}$Tata Institute of Fundamental Research, Mumbai, India}     
\centerline{$^{18}$Seoul National University, Seoul, Korea}                   
\centerline{$^{19}$CINVESTAV, Mexico City, Mexico}                            
\centerline{$^{20}$FOM-Institute NIKHEF and University of                     
                  Amsterdam/NIKHEF, Amsterdam, The Netherlands}               
\centerline{$^{21}$University of Nijmegen/NIKHEF, Nijmegen, The               
                  Netherlands}                                                
\centerline{$^{22}$Institute of Nuclear Physics, Krak\'ow, Poland}            
\centerline{$^{23}$Joint Institute for Nuclear Research, Dubna, Russia}       
\centerline{$^{24}$Institute for Theoretical and Experimental Physics,        
                   Moscow, Russia}                                            
\centerline{$^{25}$Moscow State University, Moscow, Russia}                   
\centerline{$^{26}$Institute for High Energy Physics, Protvino, Russia}       
\centerline{$^{27}$Lancaster University, Lancaster, United Kingdom}           
\centerline{$^{28}$Imperial College, London, United Kingdom}                  
\centerline{$^{29}$University of Arizona, Tucson, Arizona 85721}              
\centerline{$^{30}$Lawrence Berkeley National Laboratory and University of    
                  California, Berkeley, California 94720}                     
\centerline{$^{31}$University of California, Davis, California 95616}         
\centerline{$^{32}$California State University, Fresno, California 93740}     
\centerline{$^{33}$University of California, Irvine, California 92697}        
\centerline{$^{34}$University of California, Riverside, California 92521}     
\centerline{$^{35}$Florida State University, Tallahassee, Florida 32306}      
\centerline{$^{36}$Fermi National Accelerator Laboratory, Batavia,            
                   Illinois 60510}                                            
\centerline{$^{37}$University of Illinois at Chicago, Chicago,                
                   Illinois 60607}                                            
\centerline{$^{38}$Northern Illinois University, DeKalb, Illinois 60115}      
\centerline{$^{39}$Northwestern University, Evanston, Illinois 60208}         
\centerline{$^{40}$Indiana University, Bloomington, Indiana 47405}            
\centerline{$^{41}$University of Notre Dame, Notre Dame, Indiana 46556}       
\centerline{$^{42}$Iowa State University, Ames, Iowa 50011}                   
\centerline{$^{43}$University of Kansas, Lawrence, Kansas 66045}              
\centerline{$^{44}$Kansas State University, Manhattan, Kansas 66506}          
\centerline{$^{45}$Louisiana Tech University, Ruston, Louisiana 71272}        
\centerline{$^{46}$University of Maryland, College Park, Maryland 20742}      
\centerline{$^{47}$Boston University, Boston, Massachusetts 02215}            
\centerline{$^{48}$Northeastern University, Boston, Massachusetts 02115}      
\centerline{$^{49}$University of Michigan, Ann Arbor, Michigan 48109}         
\centerline{$^{50}$Michigan State University, East Lansing, Michigan 48824}   
\centerline{$^{51}$University of Nebraska, Lincoln, Nebraska 68588}           
\centerline{$^{52}$Columbia University, New York, New York 10027}             
\centerline{$^{53}$University of Rochester, Rochester, New York 14627}        
\centerline{$^{54}$State University of New York, Stony Brook,                 
                   New York 11794}                                            
\centerline{$^{55}$Brookhaven National Laboratory, Upton, New York 11973}     
\centerline{$^{56}$Langston University, Langston, Oklahoma 73050}             
\centerline{$^{57}$University of Oklahoma, Norman, Oklahoma 73019}            
\centerline{$^{58}$Brown University, Providence, Rhode Island 02912}          
\centerline{$^{59}$University of Texas, Arlington, Texas 76019}               
\centerline{$^{60}$Texas A\&M University, College Station, Texas 77843}       
\centerline{$^{61}$Rice University, Houston, Texas 77005}                     
\centerline{$^{62}$University of Virginia, Charlottesville, Virginia 22901}   
\centerline{$^{63}$University of Washington, Seattle, Washington 98195}       
}                                                                             
  
%
\date{\today}
\maketitle
\begin{abstract} 
 We present results of a search for R-parity-violating  
 decay of the neutralino ${\tilde{\chi}}_{1}^{0}$, taken to be the Lightest 
 Supersymmetric Particle. It is assumed that this decay 
 proceeds through one of the lepton-number violating 
 couplings ${\lambda}^{'}_{2jk}$ ($j=1,2$; $k=1,2,3$),
 and that R-parity is conserved in all other production and decay
 processes in the event. This scenario  
 provides two muons and four jets in the final state. This search is based on
 77.5 $\pm$ 3.9 pb$^{-1}$ of data, collected by the \D0 experiment at the
 Fermilab Tevatron in 
 $p\overline{p}$ collisions at $\sqrt{s}$=1.8 TeV in 1992--1995.   
 Background expected from standard model processes 
 amounts to 0.18 $\pm$ 0.03 $\pm$ 0.02 events. In the absence of 
 candidate events, the result is interpreted in terms of   
 limits on squark and gluino masses
 within the framework of the minimal low-energy supergravity
 supersymmetry model. 
  
\end{abstract}
\begin{figure}[htp]
\epsfxsize=3.6in
\epsfbox{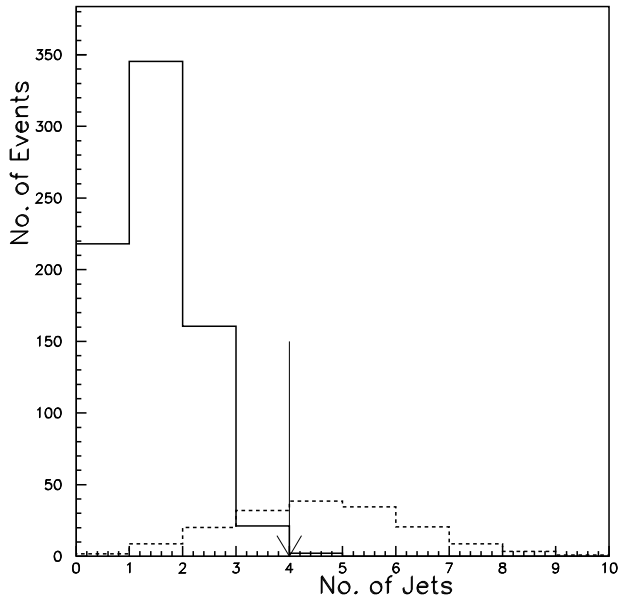}
\caption{Distribution of the number of jets per event at a typical signal
 point ($m_0$=140 GeV/$c^2$, $m_{1/2}$=90 GeV/$c^{2}$) 
(dashed line) and background sample, 
 $Z (\rightarrow \mu\mu)$ $+$ jets (solid line) for 77.5 $pb^{-1}$ integrated 
 luminosity. The vertical arrow indicates the position of the applied cut.}
\label{fig:sig}
\end{figure}

\begin{figure}[ht]
\epsfxsize=3.6in
\epsfbox{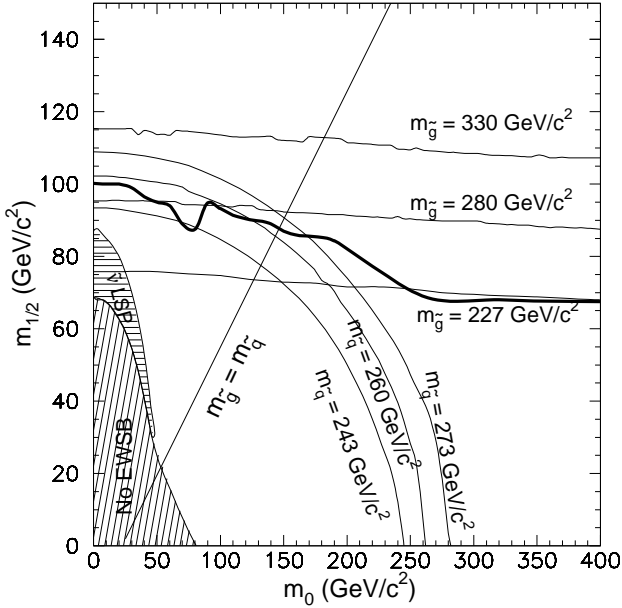}
\caption{Exclusion contour in the $(m_0,m_{1/2})$ plane for tan$\beta =2$,
$\mu\ < 0$, $A_0=0$, and finite ${\lambda}^{'}_{2jk}$ ($j=1,2$; $k=1,2,3$)
 coupling. The region below the bold line is excluded at the 95\% C.L. 
The cross 
 hatched region is excluded for theoretical reasons (see text). 
 $m_{\tilde{q}}$ and $m_{\tilde{g}}$ denote the squark and gluino masses,
 respectively.}
\label{fig:tanb2}
\end{figure}

\begin{figure}[t]
\epsfxsize=3.6in
\epsfbox{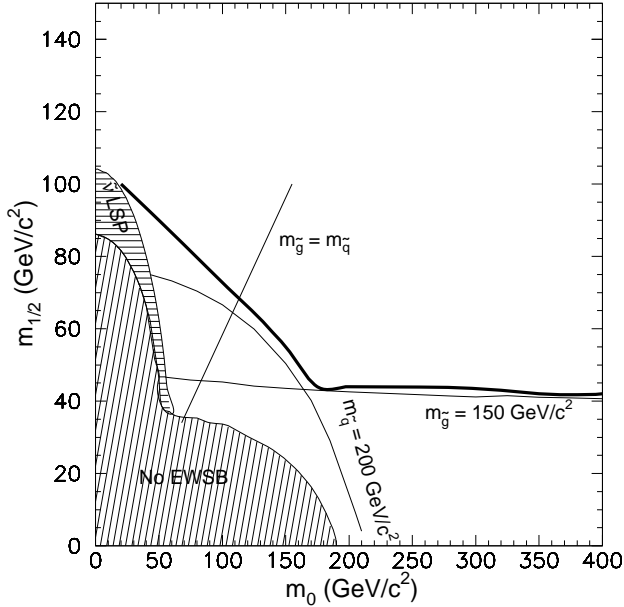}
\caption{Exclusion contour in the $(m_{0}, m_{1/2})$ plane for 
tan$\beta =6$, $\mu\ < 0$, $A_0=0$, and finite 
${\lambda}^{'}_{2jk}$ ($j=1,2$; $k=1,2,3$) coupling.}
\label{fig:tanb6}
\end{figure}

  
 A search for events with multiple leptons and {\mbox jets} is an effective
 way to look for new physics because such events do not suffer from 
 large standard model (SM) backgrounds. These events can provide evidence 
 of {\mbox R-parity-violating} (RPV) decays of supersymmetric (SUSY)
 particles~\cite{susy,fayet}. R-parity is a 
 discrete multiplicative quantum number that distinguishes 
 SM particles from their SUSY partners. It is defined 
 as R $= (-1)^{3B+L+2S}$, where B, L, and S are
 the baryon, lepton, and spin quantum numbers, respectively. R is $+1$ for SM
 particles and $-1$ for the corresponding SUSY particles. Originally, 
conservation of R-parity was imposed on supersymmetric 
  theories because the combination of lepton-number
 and baryon-number violating
 couplings in the Lagrangian could have generated several rare or forbidden
 processes at {\mbox unacceptably} high rates. One such example is the decay 
 of the proton. However, rapid proton decay as well as other rare decays
 can be prevented by not allowing  simultaneous violations of baryon and 
 lepton numbers. Thus, a small violation of R-parity cannot be excluded.\\
\indent The Yukawa coupling terms 
 in the superpotential that induce R-parity violation are:
\begin{center}
                ${\lambda}_{ijk}{L_i}{L_j}\overline{E}_k$ +
 ${{\lambda}_{ijk}^{'}}{L_i}{Q_j}\overline{D}_k$ +  
 ${{\lambda}_{ijk}^{''}}{\overline{U}_i}{\overline{D}_j}{\overline{D}_k}$,\\
\end{center}
 where $L$ and $Q$ are the SU(2)-doublet lepton and quark superfields; 
 $E$, $U$, and $D$ are the singlet lepton, up-type quark, and down-type quark 
 superfields, 
 respectively; and $i$, $j$, and $k$ are the generation indices. 
 Since $\lambda$ and $\lambda^{''}$ are antisymmetric in the first two 
 and last two indices, respectively, there
 are in total 45 possible couplings. For experimental searches it is usually 
 {\mbox assumed} that only one of the 45 couplings is non-zero. 
Since experimental 
upper {\mbox bounds} on these couplings from low-energy measurements are quite 
 stringent~\cite{bounds}, it is further assumed that 
 R-parity violation {\mbox manifests} itself only in the decay of the lightest 
 supersymmetric particle (LSP). At the same time, these couplings are
 {\mbox assumed} to be strong enough so that the LSP is unstable and decays 
 within the detector, close to the interaction vertex, which sets the scale
 for $\lambda$ at $\approx {10}^{-3}$. 
 A previous study at \D0~\cite{nirmalya} in the dielectron + {\mbox jets} 
channel, searched for such
 a decay for non-vanishing ${\lambda}_{1jk}^{'}$ ($j=1,2$ and $k=1,2,3$) 
 couplings in the framework of the minimal low-energy supergravity
 supersymmetry model ({\mbox mSUGRA})~\cite{sugra}, with
 ${\tilde{\chi}}_{1}^{0}$ as the LSP. This model
 contains five parameters: a common mass for scalars ($m_0$), 
a common mass for 
 gauginos ($m_{1/2}$), a common trilinear coupling ($A_0$, specified
 at the grand unification scale), the ratio of the vacuum expectation values
 of the two Higgs doublets (tan$\beta$), and the sign of the 
 Higgsino mass parameter ($\mu$). The LSP decay to a charged lepton and two 
 quark jets involving one of the ${\lambda}_{ijk}^{'}$ couplings
 is a viable mode for searching for SUSY at the Tevatron for the 
following reasons. The LSP can be 
 produced either directly or through cascade decays from {\mbox squarks} or 
gluinos and can subsequently decay into a lepton and 
{\mbox two} {\mbox quarks}. The
 branching fraction of this decay depends 
 on the composition of the LSP, which
 in turn depends on the mSUGRA parameters described earlier.
 Studies have shown that at the energy of the Tevatron, the 
 amount of signal in any of the lepton + jets decay channels of the LSP
 can be substantial for a large
 range of values of the mSUGRA parameters~\cite{formula,rpvdecay}. 
 Also, such events will not contain any missing energy, 
 thus making it easier to search for a RPV signal.  
 We report a study similar to the previous one~\cite{nirmalya}, for finite 
 ${\lambda}_{222}^{'}$ coupling (the study is equally valid for all the
 ${\lambda}_{2jk}^{'}$ couplings with $j=1,2$ and $k=1,2,3$), based on a
 signature of {\mbox two} energetic muons and four energetic jets. There are 
 several standard model processes that mimic this signature, e.g.,
 ${\gamma}^{*}$/$Z \rightarrow \mu\mu$, 
 $Z \rightarrow \tau \tau \rightarrow \mu \mu$, 
 $t \overline{t}\rightarrow\mu\mu$, $WW \rightarrow \mu\mu$, and accompanying
 jets.\\
\indent The \D0 detector has been described elsewhere~\cite{det}.
 The most important parts for this analysis are the 
uranium/liquid-argon calorimeter
 and the muon system. A cone {\mbox algorithm} with a cone 
 radius of 0.5 in the {\mbox {$\eta$-$\phi$}} space, where $\eta$ is the 
 pseudorapidity and $\phi$ is the azimuthal 
 angle, is used for jet identification~\cite{cone}. 
 {\mbox Muons} are identified as tracks
 that leave minimum ionizing energy in the calorimeter, and are 
 reconstructed in the muon system. An integrated luminosity of 
 77.5 $\pm$ 3.9 pb$^{-1}$ collected
 with the \D0 detector during the 1992--1995 Tevatron run at 
 $\sqrt{s}$ = 1.8 TeV is used for this analysis. The data are required
 to satisfy a trigger demanding one muon ($p_T >$ 10 GeV/$c$, $|\eta| <$ 1.7),
 and one jet ($E_T >$ 15 GeV, $|\eta| <$ 2.5). 
In the offline analysis, an event is selected only if 
  it has at least two muons within $|\eta| <$ 1.7 ($p_T >$ 15 GeV/$c$ for the 
first muon, and $p_T >$ 10 GeV/$c$ for the second muon), and at least four 
jets within $|\eta| <$ 2.5 and with $E_T >$ 15 GeV. 
The muons and jets are required to satisfy standard \D0 selection criteria 
~\cite{muonid,aplan}. 
 The muons are also required to be isolated
 from jets by a distance $> 0.5$ in the $\eta$-$\phi$ plane (this rejects
 muons coming from heavy-flavor decays, pions decaying in flight, and 
 pion-induced punchthroughs).  In addition, several other criteria
 are imposed to minimize 
 background. The {\mbox aplanarity}~\cite{aplan} of the jets in each
 event is required 
 to be greater than 0.03. The invariant mass of the two \mbox{muons} is
 required to be greater than 
 5 GeV/$c^2$, which helps to reject low-energy resonances 
(e.g. $J/{\psi}$) and 
 spurious combinations of muon tracks. $H_T$, the scalar sum of 
 $E_T$ of all muons and jets that pass kinematic and
 fiducial requirements, is required to be greater than 150 GeV. \\
\indent Of 
 the original 230,688 events passing the trigger requirements, none survive
 the above selections. The expected backgrounds from the two main SM  
 channels, $Z (\rightarrow \mu \mu)$ $+$ jets and 
 $t \overline{t} (\rightarrow \mu \mu)$ $+$ jets, 
 are shown in Table~\ref{tab:backgrd},
 along with their statistical (first) and systematic (second) uncertainties. 
 The contribution to background events from 
  $Z$ production is estimated from a sample of 21,000 $Z$ + jets events,
 generated using {\sc vecbos}~\cite{vecbos}. 
 A total of 254,000 $t \overline{t}$ events, generated with
 {\sc herwig}~\cite{herwig}, are used to estimate the contribution 
 from this background. The \D0 detector is simulated using 
 a {\sc geant}-based package~\cite{geant}, which provides efficiencies 
of the {\mbox selection} criteria for signal and background events. 
  We illustrate in Fig.~\ref{fig:sig}, the effect of one of the
 selection criteria (number of jets in an {\mbox event}) on events 
 at a typical signal 
 point ($m_0$=140 GeV/$c^2$, $m_{1/2}$=90 GeV/$c^2$, $A_0$=0, tan$\beta$=2,
 $\mu<0$) and on events from the background channel $Z (\rightarrow \mu \mu$)
 $+$ jets. 
 The arrow in Fig.~\ref{fig:sig} indicates the minimum number of jets in 
accepted events. 
\begin{table}[h!]
\caption{Summary of major backgrounds. First error is statistical and
second error is systematic.}
\begin{center}
\begin{tabular}{l l} 
Background   & Expected events for 77.5~pb$^{-1}$ \\
process    &                                      \\ \hline
$Z (\rightarrow \mu \mu)$ $+$ jets     & $0.140\pm0.031 \pm 0.015 $ \\ 
$t \overline{t} (\rightarrow \mu \mu)$ $+$ jets  
    & $ 0.042 \pm 0.002 \pm 0.013 $ \\ 
Total                               & $0.182 \pm 0.031 \pm 0.020$  
\end{tabular}
\label{tab:backgrd}
\end{center}
\end{table}
The instrumental background, which arises from misidentification
 of jets as muons, is negligible in this analysis. As can be seen from 
 Table~\ref{tab:backgrd}, the expected number of background events is 
 quite small. The statistical error arises from a combination of fluctuations 
 in the Monte Carlo events and uncertainties in the muon and jet 
 identification efficiencies. The systematic error arises due to uncertainty 
 in the jet energy scale and in the values of production cross sections.\\
\begin{table}
\begin{center}
\caption{Efficiency ($\epsilon$) multiplied by branching fraction ($B$), and
 expected event yield $\langle N \rangle$, for several points in the
($m_0$, $m_{1/2}$) 
parameter space (for tan$\beta$=2, $A_0$=0, and $\mu<0$).}
\begin{tabular}{c c c c} 
$m_0$ (GeV/$c^2$)  & $m_{1/2}$ (GeV/$c^2$) &  $\epsilon  B (\%)$  
& $\langle N \rangle$\\\hline
   0 & 100&$0.60 \pm0.07^{+ 0.05 } _ {-0.03}$&$  3.0\pm 0.4$ \\
  80 & 90&$0.74 \pm0.08^{+ 0.06 } _ {-0.04}$&$  2.7\pm 0.3$ \\
 80 & 110&$0.34 \pm0.04^{+ 0.03 } _ {-0.03}$&$  0.6\pm 0.1$ \\
 190 & 90&$ 0.78 \pm0.06^{+ 0.05 } _ {-0.03}$&$  2.1\pm 0.2$ \\
 260 &  70&$ 0.42 \pm0.04^{+ 0.03 } _ {-0.02}$&$  2.7\pm 0.3$ \\
 400 &  90&$0.31 \pm0.04^{+ 0.02 } _ {-0.02}$&$  0.8\pm 0.1$ 
\end{tabular}
\label{tab:signal}
\end{center}
\end{table}
\indent Signal events are generated with {\sc isajet}~\cite{isajet}, 
modified to 
 incorporate RPV decays based on the formalism of Ref.~\cite{formula}. 
  For each signal sample, 
 the value of efficiency multiplied by the branching fraction of 
 $p\overline{p}\rightarrow$ $\ge$ two muons and $\ge$ four jets 
 is estimated in the same
 way as described above for the SM background. Table~\ref{tab:signal}
 shows these values and the event yields 
 expected from an integrated luminosity of
 77.5 pb$^{-1}$ for several points in the ($m_0$, $m_{1/2}$) parameter 
 space.\\
\indent Since the expected SM background is compatible with absence of 
 observed events, we proceed to 
 determine the region in mSUGRA space that can be excluded.
An upper limit  at the 95\% confidence level (C.L.) on the cross section 
for signal is obtained for each point in the ($m_0,m_{1/2}$)
 plane for fixed values of $A_0$=0, {\mbox {$\mu <$ 0}},
 and tan$\beta$ = 2 and 6. 
 A technique based on Bayesian statistics~\cite{bayes} is used for 
 this purpose, with a flat 
 prior for the signal cross section and Gaussian priors for 
 luminosity, efficiency, and expected background. The limits on the 
 measured cross section are then compared with the leading-order SUSY
 prediction given by {\sc isajet}, to find an excluded region in the 
 ($m_0,m_{1/2}$) plane. Figs.~\ref{fig:tanb2} and~\ref{fig:tanb6} 
 show the regions of parameter space (below the bold lines) excluded at the 
 95\% C.L. for tan$\beta$ = 2 and 6, respectively.\\
\indent The shaded areas in the left-hand corners of the figures 
 indicate the regions where either the model
 does not produce electroweak symmetry breaking or the 
 lightest neutralino is not the LSP. The area in the ($m_0$, $m_{1/2}$)
 plane excluded by experimental searches at LEP~\cite{lep} already extends  
 beyond the shaded areas.
The exclusion contour in Fig.~\ref{fig:tanb2} follows essentially a 
contour of {\mbox constant} squark mass 
 ($m_{\tilde{q}}$ = 260 GeV/$c^2$) for low $m_0$ values. This is because
 pair production of squarks is the {\mbox dominant} SUSY process that 
 contributes to the signal in that region. Production
 of gluinos, ${\tilde{\chi}}_2^0$, and ${\tilde{\chi}}_1^0$ becomes 
 dominant at larger 
 values of $m_0$, where the masses and production cross sections of these 
 particles are approximately independent of $m_0$. 
 The exclusion contour therefore becomes approximately 
 independent of $m_0$ for $m_0 >$ 250 GeV/${c^2}$.\\
\indent The value of $A_0$ does not affect the results significantly,
  since it changes
 only the third generation sparticle masses. Both for $\mu>$ 0, and for higher
 values of tan$\beta$ (see {\mbox Fig. 3} for the exclusion contour at 
tan$\beta$ = 6),
 the sensitivity of this search diminishes, because of
 the change in the composition of the LSP, which leads to a decrease of the  
 branching fraction of the LSP into muons~\cite{rpvdecay}.\\
\indent In conclusion, we have searched for RPV decay of the neutralino
 ${\tilde{\chi}}_1^0$ into a muon and two jets in 77.5 pb$^{-1}$ of data. 
 No candidate events were found.
 This result is presented as an exclusion contour in the 
{\mbox mSUGRA} 
 ($m_0$, $m_{1/2}$) parameter space for $A_0$=0, tan$\beta$=2 and 6, and
 $\mu <0$. In particular, for tan$\beta$ = 2, squark
 masses below 240 GeV/$c^2$ (for all gluino masses) and gluino masses below
 224 GeV/$c^2$ (for all squark masses) can be excluded. For equal 
 masses of squarks and gluinos the mass limit is 
 265 GeV/$c^2$.\\
\indent We thank the staffs at Fermilab and collaborating institutions, 
and acknowledge support from the 
Department of Energy and National Science Foundation (USA),  
Commissariat  \` a L'Energie Atomique and 
CNRS/Institut National de Physique Nucl\'eaire et 
de Physique des Particules (France), 
Ministry for Science and Technology and Ministry for Atomic 
   Energy (Russia),
CAPES and CNPq (Brazil),
Departments of Atomic Energy and Science and {\mbox Education} (India),
Colciencias {\mbox (Colombia),}
CONACyT (Mexico),
Ministry of Education and {\mbox KOSEF} (Korea),
CONICET and UBACyT (Argentina),
The Foundation for Fundamental Research on Matter (The {\mbox Netherlands)},
PPARC (United Kingdom),
Ministry of {\mbox Education} (Czech Republic),
and the A.P.~Sloan Foundation.

\end{document}